\documentstyle[12pt,aasms4]{article}

\newcommand{\etal}{{\it et~al.\ }}
\newcommand{\ie}{{\it i.e.,~}}
\newcommand{\eg}{{\it e.g.,~}}

\lefthead{Ferrara, Bianchi, Dettmar, \& Giovanardi}
\righthead{Scattering by Dust and Emission Line Ratios}

\begin{document}

\title{The Effect of Light Scattering by Dust\\
in Galactic Halos on Emission Line Ratios}

\author{Andrea Ferrara}
\affil{ Osservatorio Astrofisico di Arcetri\\
Largo E. Fermi, 5, 50125 Firenze, Italy}
\authoraddr{ Largo E. Fermi, 5, 50125 Firenze, Italy}
\authoremail{ferrara@arcetri.astro.it}

\author{ Simone Bianchi}
\affil{ Univ. di Firenze, Dipartimento di Astronomia e Scienza dello Spazio\\
Largo E. Fermi, 5, 50125 Firenze, Italy}

\author{Ralf-J\"urgen Dettmar}
\affil{ Astronomisches Institut, Ruhr-Universit\"at Bochum\\ D-44780
Bochum, Germany}

\and 

\author{Carlo  Giovanardi}
\affil{ Osservatorio Astrofisico di Arcetri\\
Largo E. Fermi, 5, 50125 Firenze, Italy}

\begin{abstract}
We present results from Monte Carlo simulations describing the radiation
transfer of $H\alpha$ line emission, produced both by HII regions in the
disk  and in the diffuse ionized gas (DIG), through the dust layer of the
galaxy NGC891.
This allows us to calculate the amount of light originating in the HII
regions of the disk and scattered by dust at high $z$, and compare it
with the emission produced by recombinations in the DIG.   
The cuts of photometric and polarimetric maps along the $z$-axis
show that scattered light from HII regions is still 10\% of
that of the DIG at $z\sim 600$~pc, whereas the  
the degree of  linear polarization is small ($<1$\%). 
The importance of these results for the determination of intrinsic
emission line ratios is emphasized, and the significance and possible 
implications of dust at high $z$ are  discussed.
\end{abstract}

\keywords{dust, extinction---ISM: abundances---galaxies: spiral---HII:
regions---polarization---radiative transfer---scattering}


\section{Introduction }
The thick ($z\sim 1 $~kpc) {\it Reynolds layer} of diffuse ionized gas (DIG)
discovered in the Galaxy (Reynolds 1990) poses some of the most challenging 
problems for our understanding
of the large scale structure of the Galactic ISM. Not only its vertical support against
the gravitational pull of the Galaxy, but also the heating/ionization sources, and the
excitation conditions are not well understood and they are matter of current debate.
In addition, one would like to understand its possible relationship with the stellar activity
in the disk, \ie HII regions, supernova remnants, superstructures etc.
 
Such ionized layers are  also found in several edge-on spiral galaxies, thus lending
support to the perception that this structure is not peculiar to the Milky Way.
In addition, edge-on galaxies provide a  direct way to study the high $z$
gas since disk and halo separate in projection. The Sbc galaxy NGC891
has to be considered by far the best candidate for this type of studies both because
of the huge wealth of data available, and for its close similarity to the Milky Way
(van der Kruit 1984).
 
The easiest way to study the DIG is via deep H$\alpha$-images (Rand \etal 1990,
Dettmar 1990); these studies have shown not only the large vertical extension
of the gas but also several  small scale structures (asymmetries, filaments, ``froth'')  and a
conspicuous number of dust lanes. However, even more relevant to the unsolved problems
mentioned above might be the spectroscopic observations of the DIG. Dettmar \& Schulz
(1992) derived the following ratios for the detected
emission lines in the range $z=0-900$~pc: $[NII](\lambda 6583$\AA$)/H\alpha \sim 0.32-1.1$,
$[SII](\lambda 6716$\AA$)/H\alpha \sim 0.11-0.6$. These values at the midplane are close
to the analogous ones observed in HII regions ($\sim 0.3, 0.08$, respectively),
but they depart considerably at higher $z$. Similar changes have recently
been reported for the DIG in NGC4631 (Golla \etal 1996). This evidence suggests
different
excitation conditions, possibly due to a change in spectrum the radiation field; 
it also
poses a relevant question
concerning the amount of light originating in the disk (where the most obvious 
ionization
sources are located) and light scattered back from dust at high $z$, which we address
in this work. In principle, it is also expected that the scattered light is linearly
polarized and thus
it is useful to produce synthetic polarization maps as well as photometric maps to be compared with the data.
As a by-product, one hopes that this kind of studies can have a positive feedback
on searches for dust in the halo, whose presence is predicted on several 
grounds (Ferrara \etal 1991, Ferrara
1993), and recently marginally detected  by Zaritsky (1994), which could have important
cosmological implications.
 
To answer to the above questions we have performed extensive Monte Carlo computations
simulating NGC891 in detail, which are presented in the next Section. The results are
discussed in Section 3 and the conclusions are drawn in Section 4.
 
\section{Monte Carlo simulations: setup}
We want to follow the radiation tranfer of H$\alpha$ photons, produced both by
HII regions in the disk and in the DIG, through the
dust layer of the galaxy NGC891 as a function of height,  $z$.  To this aim we
use a Monte Carlo code simulating dusty spiral galaxies, modelled as bulge+disk
systems immersed in a given spatial distribution of dust grains. The extinction parameters
(absorption and scattering) of dust grains are calculated from Mie's theory for a
(usually MRN) distribution of grain sizes and materials; the radiation transfer is carried
out for the four Stokes parameters, thus enabling us to produce photometric and
polarimetric maps of the galaxy, once the central optical depth, inclination and bulge-to-total
ratio are (\ie Hubble type) given. A complete description of the code can be found in Bianchi, Ferrara
\& Giovanardi (1996).
 
To quantify the contribution of  H$\alpha$ photons produced by the HII regions
located in the disk relative to the ones truly produced by recombination events in
the extraplanar DIG, we have run the code in the following configuration.
We model the dust disk according to Kylafis \& Bahcall (1987, KB), who concluded 
that a good representation of the dust distribution  has a sech$^2$ functional form
in the vertical direction (we have tried also exponential and gaussian
distributions, see Sec. 3)
and is exponential in the horizontal (\ie in the plane of  NGC891) direction 
with scale lengths
$z_d = 0.22$~kpc and $r_d = 4.9$~kpc, respectively. The central optical depth 
at $\lambda=6563$\AA\  measured along the symmetry axis, also deduced 
from KB, is $\tau = 0.445$.
The value of the inclination adopted is $i=88.5^o$ (van der Kruit 1984).
Next, we specify the distribution of the DIG. This is taken from Dettmar (1990) who
concludes that the average H$\alpha$ distribution (as derived from the emission
measure; we recall that the electron scale height is twice the scale height of the
emission measure) of the DIG in
NGC891 is very well approximated by an exponential distribution with a scale height
$z_{DIG}=0.3-0.5 kpc$; in our simulations we have used the mean value $0.4$~kpc. 
Of course, one has to
keep in mind that this average distribution does not take into account the numerous
irregularities observed. The ionized layer is assumed to be uniform in the radial direction.
As a final step it is necessary to assume a distribution for the HII regions. 
Obviously, the true distribution
of the HII regions of NGC891 is not known  due to the heavy obscuration occurring
in the intervening equatorial dust lane.
Our strategy has then been to exploit the fact that NGC891 is considered to be almost
a ``twin'' of the Milky Way and to assume that the radial distribution of HII regions is
therefore the same as in the Galaxy for which an extensive literature can be found. 
We rely on the results of Lockman (1990),  who presents in his Fig. 2 the surface density
of all radio HII regions at galactocentric radius $\varpi > 1.5$~kpc and Galactic latitude
$\vert \ell \vert > 10^o$. These results represent the most complete survey of Galactic
HII regions to our knowledge; we assume that, properly scaled, the {\it shape} of the
radial distribution is the same in NGC891. As for the (vertical) $z$-distribution of HII regions,
although its precise form is not firmly determined,  the scale height has been estimated
in the Galaxy to be $\sigma \sim 70$~pc (Manchester \& Taylor 1981; Reynolds 1989). Dettmar
(1990) finds for the HII regions in NGC891 a gaussian distribution with $\sigma = 170$~pc;
we have therefore used this value. However, as long as $\sigma \ll z_{DIG}$ the
results are quite insensitive to the precise value of $\sigma$.
 
We are left with the problem of specifying the values  of  the  H$\alpha$
luminosity of the DIG, $H\alpha(DIG)$, and of the HII regions, $H\alpha(HII)$.
The total H$\alpha$ luminosity is given by the following sum:
$$H\alpha(TOT) = H\alpha(HII)+ H\alpha(DIG);\eqno(1)$$
we can express the $H\alpha$ luminosity of the HII regions and of the DIG by means of the 
{\it observed} luminosity, via the following equations
$$ H\alpha^{(o)}(HII)=p H\alpha(HII),\eqno(2) $$
and
$$ H\alpha^{(o)}(DIG)=q H\alpha(DIG),\eqno(3) $$
where $H\alpha^{(o)}(HII)$ and $H\alpha^{(o)}(DIG)$ denote $H\alpha$ photons
produced by HII regions, and by the DIG, respectively, escaping in the line 
of sight either directly or after being scattered.
We determine $p$ and $q$ from simulations of HII regions or DIG alone immersed in the dust
disk counting the number of photons that have been scattered in the line of sight;
from our simulation we derive $p=0.54$ and $q=0.65$.

The final step is to fix the ratio $x= H\alpha(HII)/H\alpha(TOT)$; this can be done only
on a statistical basis. Kennicut \etal (1989) have measured $H\alpha$ fluxes from
the detected HII regions in 30 nearby spirals and Irr galaxies. From their Table 1,
corrected for the faint HII regions according to their Table 3, the total average HII
luminosity  for the 13 Sb-Sc galaxies (encompassing the Hubble type of NGC891)
in the sample comes out $\langle H\alpha(HII)\rangle = 9\times
10^{40}$~erg~s$^{-1}$. Since the total $H\alpha$
luminosity derived by Dettmar (1990) is  $H\alpha(TOT)=1.4\times 10^{41}$~ergs~s$^{-1}$,
it follows that the best estimate is $x=0.64$. 
This value is consistent with the most recent estimates for spiral galaxies:
Walterbos \& Braun (1994) found $x=0.6$ for M31; Veilleux \etal estimated
$x=0.7$ for NGC3079; finally Wyse \etal (1995) obtain $x=0.47$ and $x=0.59$
for the galaxies NGC247 and NGC7793, respectively. 
Given the uncertainties associated with this value, we show results for three	
cases, namely, $x=0.5$, $x=0.6$ and $x=0.7$ that encompass the experimental 
limits.
 
Our simulations calculate the radiation transfer for  the HII regions and the DIG in the dust
disk separately and we then co-add the two final images scaling their intensities appropriately.
If $N$ is the number of photons (typically $N=3\times 10^7$) for each configuration  
in the optically transparent case, then we scale by $Nq(1-x)/x$ the number of photons
for the DIG map and  by $Np$ the number of photons of the HII regions map, as can 
be derived from eq. (1) substituting the above expressions for $x$, $p$ and $q$.
The spatial resolution of our final maps is 75~pc/pixel. 
 
\section{Results}
In Figures \ref{0r0} and \ref{1r0} we show the results for two cuts along the $z$-axis, the first
through the center of the galaxy ($\varpi=0$), the second at $\varpi=4.9$ kpc,
corresponding to one scale length of the dust disk.
In each figure we show separately, for three selected values of $x$, the
$H\alpha$ luminosity profiles of the two components, HII regions and the DIG,
the ratio between them and the linear polarization profile of the $H\alpha$ for 
the complete HII+DIG model.
The effect of the central hole in the known HII regions distribution is not relevant 
in both cases, since for $\varpi=0$ it affects only a region of 50 pc around
the center of the image (in the $z$ direction) due to the large
inclination of the galaxy; the cut at $\varpi=4.9$ kpc (Fig.~\ref{1r0}) is well 
outside the hole. Also, assuming an exponential or gaussian dust vertical 
distribution with the same scale height does not produce any significant change 
in the results obtained.

The two figures clearly show that the contribution by scattered $H\alpha(HII)$ 
is still 10\% of
that of the DIG at an height of $z\sim 600$ pc; 
at these heights both the profiles and the HII/DIG ratios are rather symmetric, because
the asymmetries induced by inclined dust disks can be seen only where absorption
is high, \ie  near the galactic plane (Bianchi \etal 1996).

The degree of  linear polarization is rather small ($<1$\%) mainly because the main
source of polarized radiation ($H\alpha$ photons coming from HII
regions and scattered by the extraplanar dust) is highly diluted by
the local unpolarized DIG emission.

The polarization vectors are found to be perpendicular to the plane: this is a
known effect for scattering by standard dust spheroidal particles 
(as can be seen in similar configurations of emitters and scatterers
presented in Bianchi \etal 1996).

\section{Conclusions}

The predicted (see Figs. \ref{0r0} and \ref{1r0}) large contribution of scattered light 
originating in the plane of NCG891 to the total intensity of the $H\alpha$
line emission at considerable height $z$ above the disk can be compared 
with the observed variations of line ratios with $z$. 

As discussed in the Introduction, emission line ratios of       
$[NII]/H\alpha$ or $[SII]/H\alpha$ become gradually higher away from the 
plane. From our results, this gradient can be qualitatively understood as
being produced by dilution of the intrinsic DIG ratios by light originating
in HII regions and scattered into  the line of sight by high $z$ dust grains. 

If we assume that the line ratios at the outermost position ($z\sim
20" \sim 900$~pc) reported
by Dettmar \& Schulz (1992) represents the intrinsic line ratios of DIG
one can get a rough estimate at the intermediate position ($z\sim 10"$).  
From inspection of Fig.~\ref{1r0}, the contributions  of scattered 
$H\alpha(HII)$ and $H\alpha(DIG)$ at this position is close to unity.
This results in line ratios of $[NII]/H\alpha \sim 0.7$ and $[SII]/H\alpha
\sim 0.35$, matching the observed values quite well.  

The recently reported non-detection of the $HeI~5876$\AA\ recombination line
from the DIG of the Galaxy (Reynolds \& Tufte 1995) has questioned the 
predictions from photoionization 
models and in turn stimulated extensive searches for this critical diagnostic
line in extraplanar DIG of external galaxies (Rand 1995). The significant   
contribution of scattered light predicted by the present work might have
an important effect on this measure, since out to $z\sim 600$~pc 
part of the $HeI~5876$\AA\ line emission might originate in HII regions in the disk. 
   
The level of linear polarization of the scattered light 
expected from our models is rather low, typically $< 1\%$. Therefore 
it seems currently unfeasible to separate the contributions of the
HII regions and DIG components from polarization experiments. However,
the inclusion of non-spherical grains could increase the polarization 
degree, although we do not expect a large gain since polarization here 
is mostly produced  by scattering rather than by transmission.    

One of the assumptions of our model is that the dust distribution has
a scale height of 220~pc (KB), much smaller than the DIG one. This would 
imply that the dust-to-gas ratio (D/G) must also be a function of $z$; this
is a serious concern for any model addressing the thermal balance of
the DIG via photoelectric heating by dust grains (see, \eg 
Cox \& Reynolds 1992). If, however, in the Reynolds layer the  D/G ratio
is  similar to that  in the plane, this additional dust component could
increase the scattered light contribution at high $z$.  
Such a low density dust layer might exist and  extend to several kpc;  
a viable explanation  for its origin is provided by 
the injection of dust by radiation pressure on grains above powerful OB
associations (Ferrara \& Shull 1996). Preliminary evidence for such extended  
dust halos has been already presented by Zaritsky (1994). 
We have run test cases in which the scale height of the dust is the same
as the scale height of the DIG (800~pc), keeping the value of $\tau=0.445$, 
implying a D/G ratio essentially independent of $z$ and equal to that in
the plane. The results for the HII/DIG ratios shown in Figs. 1 and 2 
are increased by a factor of $\sim 2$. Thus, emission line ratios provide 
a sensitive tool to investigate the vertical dust distribution.

\acknowledgements

We thank B. Benjamin and R. Walterbos for useful discussions, and an anonymous
referee for comments that have considerably improved the paper.
Part of this work has been supported by German/Italian Vigoni exchange program.


\figcaption[fig1.ps]{ $z$-axis cuts through the center of the galaxy ($\varpi=0$)
for (left column) $H\alpha$ luminosity profiles corresponding to contributions 
of HII regions and DIG component, (middle) ratio of the two components, and (right)
 profiles of the linear polarization degree.  Plots refer to the cases of 
$x$=0.7, 0.6, 0.5 (from top to bottom) as discussed in the text.\label{0r0}}

\figcaption[fig1.ps]{ Same as Fig. 1, but for $\varpi=4.9$ Kpc corresponding to
one scale length of the dust disk.\label{1r0}}

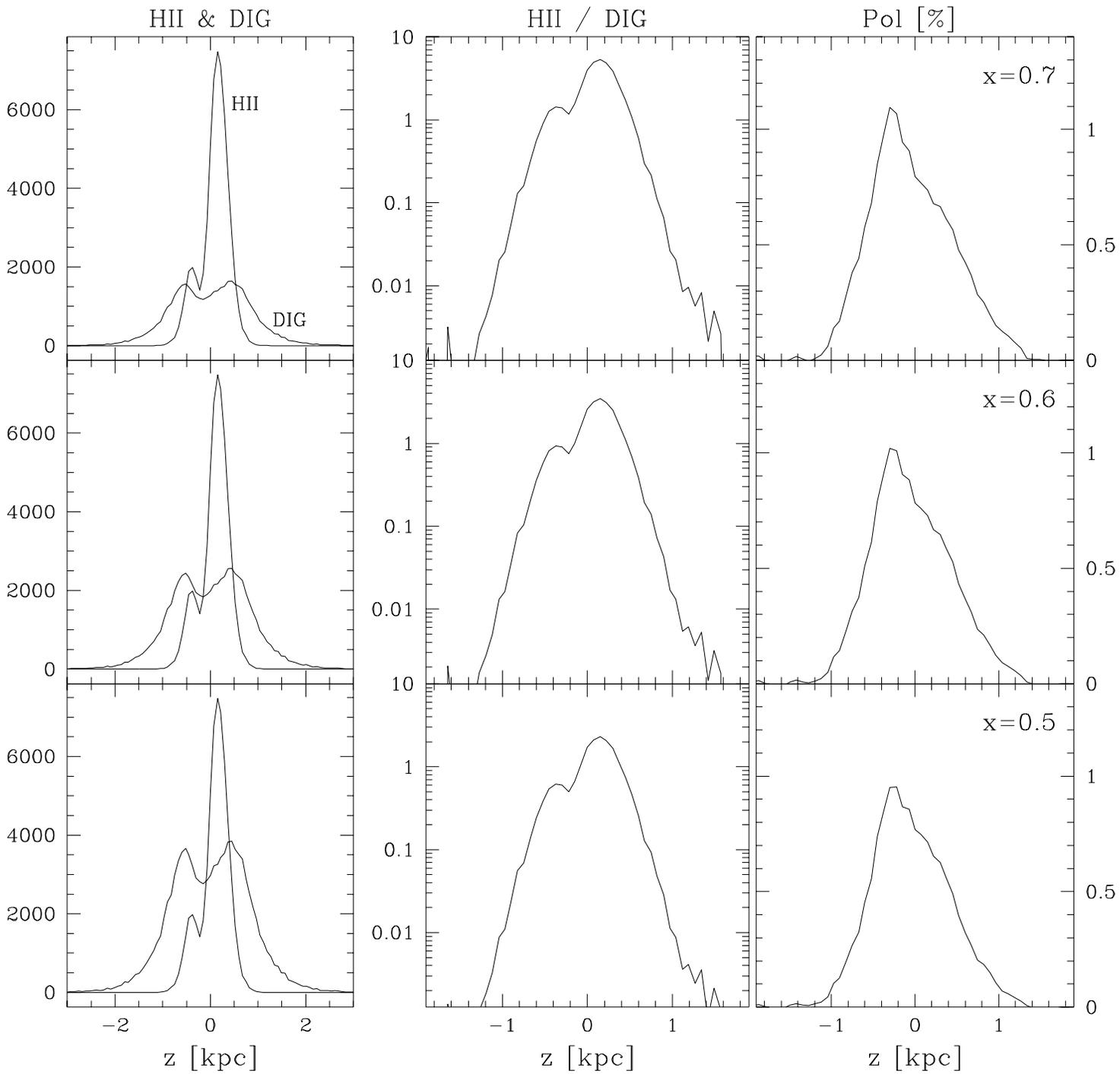
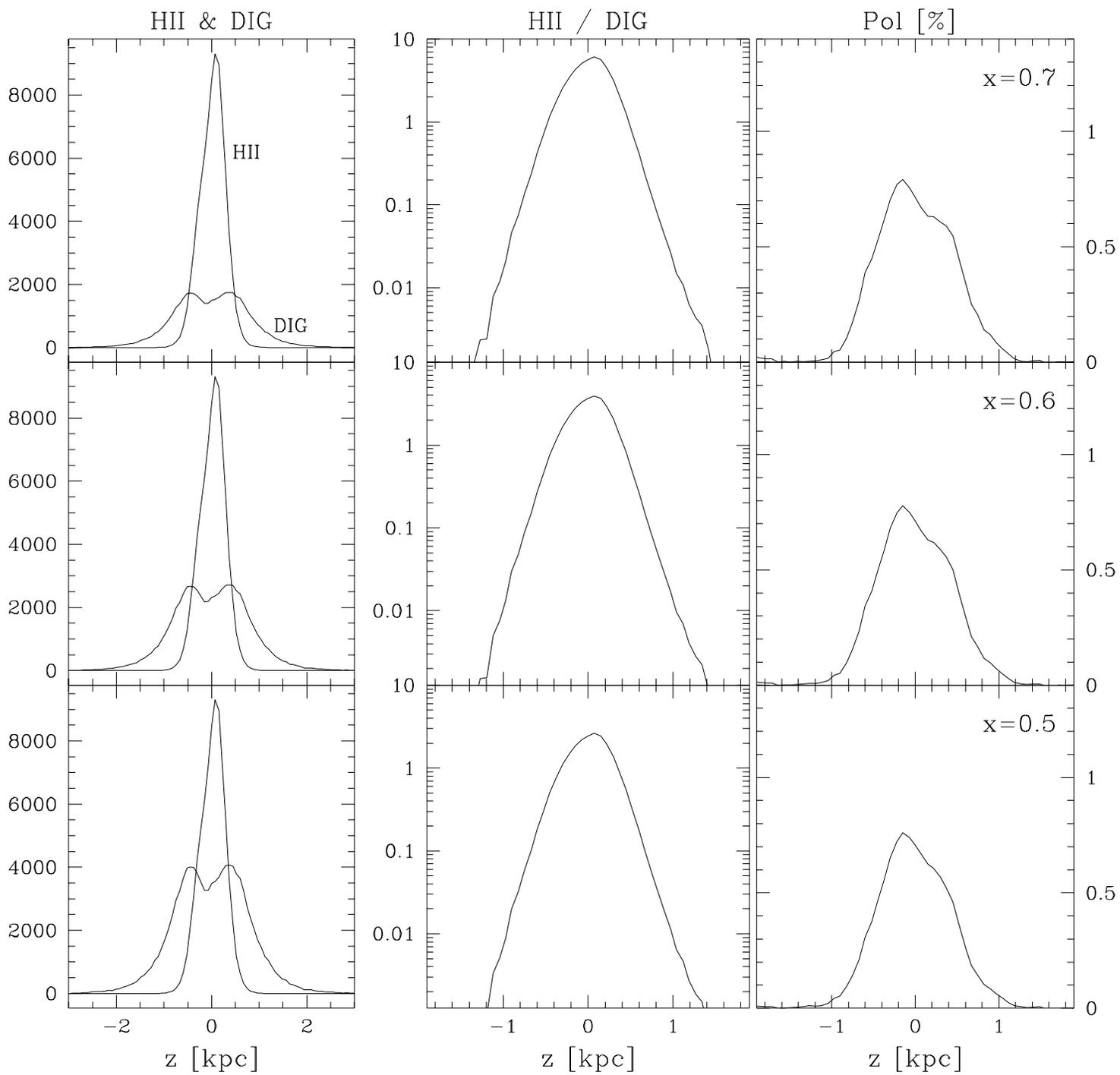
\end{document}